\documentclass[a4paper]{article}

\usepackage{multicol}
\usepackage{amsfonts}
\usepackage{color}
\usepackage{graphicx}
\usepackage{pstcol}
\usepackage[all]{xy}

\newfont{\teneufb}{eufb10 scaled 1000}
\newfont{\footeufb}{eufb10 scaled 912}

\definecolor{dark-green}{rgb}{0,0.7,0}
\definecolor{dark-blue}{rgb}{0,0.1,1}
\definecolor{red}{rgb}{0.8,0,0}
\definecolor{dark-red}{rgb}{0.5,0,0}
\definecolor{med-red}{rgb}{0.8,0.2,0.5}
\definecolor{dark-yellow}{rgb}{1,0.8,0}

\newpsobject{showgrid}{psgrid}{subgriddiv=1,griddots=10,gridlabels=6pt}

\begin{document}

\title{A Characterisation of the Weylian Structure of Space--Time by Means of Low Velocity Tests}

\author{Claus L\"ammerzahl\thanks{e-mail: claus.laemmerzahl@uni-duesseldorf.de}\\
{\normalsize Institute for Experimental Physics, Heinrich--Heine--University D\"usseldorf}\\
{\normalsize Universit\"atsstra{\ss}e 1, 40225 D\"usseldorf, Germany}}

\maketitle

\begin{abstract}
The compatibility axiom in Ehlers, Pirani and Schild's (EPS) constructive axiomatics of the space--time geometry that uses light rays and freely falling particles with high velocity, is replaced by several constructions with low velocity particles only.
For that purpose we describe in a space--time with a conformal structure and an arbitrary path structure the radial acceleration, a Coriolis acceleration and the zig--zag construction.
Each of these quantities give effects whose requirement to vanish can be taken as alternative version of the compatibility axiom of EPS.
The procedural advantage lies in the fact, that one can make null--experiments and that one only needs low velocity particles to test the compatibility axiom.
We show in addition that Perlick's standard clock can exist in a Weyl space only.
\end{abstract}

\section{Introduction}

Because of the direct physical interpretability of all introduced notions and the logical self--consistency it is most preferable to introduce the space--time geometry by means of a constructive axiomatics.
For the classical domain this task was solved in a generally accepted way by Ehlers, Pirani and Schild (1972), EPS.
They used point particles and light rays as {\it primitive objects} and axiomatized their dynamical behaviour by means of mathematical statements describing {\it basic experiences}.
For light rays the basic experience consists in the fact that there are only two light rays from any world line to a neighbouring event.
For massive point particles the equivalence principle is a basic experience.
This leads to the conformal and to the projective structure, respectively.
These two structures are not compatible in the sense that point particles obeying the equivalence principle may be non--causal with respect to the conformal structure.
The compatibility between these two structures can be achieved by demanding that no point particle is faster than light, but can chase light rays arbitrarily close.
This compatibility results in the geometrical notion of a Weylian structure which is defined by an equivalence class of metrics $[g]$ and a 1--form $a$.

A more general approach to the Weylian structure has been given by Coleman and Korte (1984) who showed that the compatibility of a path structure with the conformal structure (in the sense of local compatibilty with the Special Theory of Relativity) necessarily leads to a Weylian structure.
They proved that this is true even for higher order structures, Coleman and Korte (1993).

The above versions of compatibility rest mainly on the high velocity behaviour of massive point particles: The whole light cone has to be filled up with particle paths.
This means that one has to transfer an infinite amount of energy to the point particles in order to test the compatibility in the above form.
However, the compatibility does not show up in the high velocity behaviour of massive point particles only, but also in certain low velocity effects.
This is what we intend to do in this article.
We want to describe some low velocity effects whose requirement to vanish can be taken as alternative to the compatibility axiom using fast particles with velocities arbitrarily close to the velocity of light.
At least in the classical domain massive low velocity particles are more easy to handle with than fast particles.

In the following we very shortly review the EPS scheme for introducing the conformal and
projective structures and their compatibility axiom. Then we introduce measurable
quantities whose definitions do not depend on the notion of the projective structure.
Only light rays are needed. These quantities are the radial and Coriolis acceleration and
the rotation of a zig--zag construction. We will show that with these quantities it is
possible to replace the compatibility axiom of EPS by demanding for freely falling low
velocity particles a certain outcome for measurements of these quantities. The procedural
advantage of this form of the compatibility axiom lies in the fact that one can make
measurements with slow particles and some of these measurements are null experiments. The
only things we need are some clock, light rays and a device to register whether two light
rays propagate along the same direction.

In this connection also the role of the standard clock defined by Perlick (1987) will be interpreted further insofar as we show that such a clock can be constructed uniquely in Weylian space--times only.

\section{The conformal and the projective structure}

We assume that space--time is a differentiable manifold $\cal M$. This has also been
axiomatically founded by EPS, and, in an alternative way, in a series of papers by
Schr\"oter and Schelb (1988), (1992), and Schelb (1992).

The conformal structure of space--time is introduced by EPS by means of light rays
possessing  essentially the property that there are only two light rays between a
world--line and a nearby event (for a more rigorous formulation, see e.g. EPS or Perlick
(1997)). This and an additional topological requirement allows it to define a conformal
class of metric tensors, the {\it conformal structure}
\begin{equation}
[g] := \{g^\prime\mid g^\prime = e^\lambda g,\;\; \lambda \; \hbox{function on} \;
{\cal M}\}
\, ,
\end{equation}
where $g$ is a non--degenerate second rank tensor of signature $-2$. Two representatives
$g, g^\prime \in [g]$ are related by a conformal transformation $g^\prime = e^\lambda g$.
For each $g\in[g]$ there is a unique torsionless covariant derivative ${\buildrel {\{\}}
\over D}$ fulfilling ${\buildrel {\{\}} \over D}_u g = 0$ for all vector fields $u\in T{\cal M}$.

The tangents $l$ to the light rays are characterized by $g(l, l) = 0$, that is, light rays are null curves.
In addition, light rays obey the equation of motion
\begin{equation}
{\buildrel {\{\}} \over D}_l l \sim l \, .
\end{equation}

The second type of primitive objects, the massive point particles described by  paths $P(\tau)$, are assumed (i) to obey the weak equivalence principle (universality of free fall) and (ii) to lead to a unique determination of the path of the particle if the velocity  $v \in T{\cal M}$ (tangent to the path) is given at a certain point $p\in {\cal M}$.
Therefore, the particle paths are given by some path structure, which is an equivalence class of curves $\gamma: \mathbb{R} \rightarrow {\cal M}: \tau \mapsto x(\tau)$ which tangent vectors $v = dx(\tau)/d\tau$ an equations of the form $\displaystyle \hbox{\teneufb D}_v v := {{dv}\over{d\tau}} + H(x, v) = \alpha v$.
$H(x, v)$ is homogeneous of degree two in $v$ and can be used to define a velocity dependent connection; see e.g. Rund (1959).
The appearance of $\alpha$ is connected with the choice of the parametrisation of the path.
Coleman and Korte call $-H(x, v) + \alpha v$ the acceleration field.

Since we assume that a conformal structure is given, the equation for the path structure can formally be rewritten as
\begin{equation}
\hbox{\teneufb D}_v v = {\buildrel {\{\}} \over D}_v v + \hbox{\teneufb f}(v) = \alpha v\, . \label{pathstructure}
\end{equation}
with a chosen $g \in [g]$ and with $\hbox{\teneufb f}(v) := \hbox{\teneufb D}_v v - {\buildrel {\{\}} \over D}_v v \in T{\cal M}$.
For conformal transformations $\hbox{\teneufb f}$ transforms according to $\hbox{\teneufb f}(v) \mapsto \hbox{\teneufb f}^\prime(v) = \hbox{\teneufb f}(v) + d\lambda(v) v - {1\over 2} g(v, v) g(d\lambda, \cdot)$ (it is possible to absorb the component of $\hbox{\teneufb f}$ which is parallel to $v$ into the coefficient $\alpha$; however, this will not be done in our approach).

EPS introduced the projective structure as a special case of a path structure by means of the requirement that at each point $p$ of the manifold there is a coordinate system and a parametrisation so that for each trajectory of a massive point particle passing this point, the equation of motion is given by the inertial law $d^2 x^\mu/dt^2 = 0$.
(There are alternatives of this characterisation given by Douglas (1928), Ehlers and K\"ohler (1977), Coleman and Korte (1980), Pfister and Heilig (1991), and Coleman and Schmidt (1993).
Coleman and Korte (1987) claim to be able also to introduce the projective structure even in the presence of `non--geometrical' forces, see, however, the critique stated by Schr\"oter and Schelb (1995).)
In an arbitrary coordinate system and for an arbitrary parametrisation $\tau$ this equation reads
\begin{equation}
D_v v \sim v, \quad v := {{dx}\over{d\tau}} \, . \label{geodesic}
\end{equation}
where $D$ is some still unspecified linear connection.
By means of this equation not all of the 64 components of the connection $\langle D_{e_\mu} e_\nu, \theta^\rho \rangle$ can be measured ($e_\mu$ are a basis of $T{\cal M}$, $\theta^\mu$ its dual defined by $\langle e_\mu, \theta^\nu\rangle = \delta^\nu_\mu$).
For a chosen coordinate system and parametrisation, the connection can be measured up to projective transformations and torsion terms $S(u,v) := D_u v - D_v u - [u, v]$.
This defines an equivalence class of connections, the {\it projective structure}
\begin{equation}
[D] := \{ D^\prime \mid D^\prime = D + p\otimes\delta + \delta\otimes p + A, \; p\in
T^*{\cal M}, A\;\hbox{antisymmetric}\;{\textstyle \left(1\atop 2\right)}\hbox{-tensor
field}
\}
\end{equation}
$A$ automatically drops out of Eq.(\ref{geodesic}) and has therefore no influence on the following, that is, cannot be measured using point particles.

The geodesic equation (\ref{geodesic}) can again be reformulated by means of the Christoffel covariant derivative:
\begin{equation}
D_v v = {\buildrel {\{\}} \over D}_v v + f(v,v) = \alpha v\, . \label{geodesic2}
\end{equation}
The field of symmetric $\left({1\atop 2}\right)$--tensors $f$ is defined by $D_v v -
{\buildrel
{\{\}} \over D}_v v =: f(v, v) \in T{\cal M}$ and can be decomposed into a tracefree part
${\buildrel
{\rm o} \over f}$ and traces: $f = {\buildrel {\rm o} \over f} + {1\over 2}(\delta
\otimes p + p \otimes \delta) + {1\over 2}\widetilde a\otimes g$ with $p \in T^*{\cal M}$
and $\widetilde a \in T{\cal M}$. Inserting this decomposition into (\ref{geodesic2})
gives
\begin{equation}
{\buildrel {\{\}} \over D}_v v + {\buildrel {\rm o} \over f}(v, v) + p(v) v + {\textstyle {1\over 2}} \widetilde a g(v, v) = \alpha v \label{geodesic3}
\end{equation}
We also define the $\left({0\atop 3}\right)$--tensor by $f^\prime(u, v, w) := g(f(u, v), w)$ which in the following we denote by $f$ too.
The totally symmetric tracefree part of $f$ or ${\buildrel {\rm o} \over f}$ will be denoted by ${\buildrel {\rm s} \over f}$.

Until now we established no connection between the propagation phenomena for light and
for massive point particles. For example, the particle motion may be acausal with respect
to the metric $g$. To make these two phenomena compatible, EPS required that particles
cannot cross the light cone and that the light cone can be filled up with particle paths.
Especially this last requirement needs high energy particles to be tested. In the
classical domain it may be difficult for massive point particles to be accelerated to
velocities being arbitrarily close to the velocity of light. This {\it compatibility
condition} gives a connection between the Christoffel connection ${\buildrel {\{\}} \over
D}$ and the connection $D$: $f(v, v) = \gamma v\;\forall v$ with $g(v, v) = 0$ and for
some unspecified function $\gamma$. This requirement results in $D \in [{\buildrel W
\over D}]$ where ${\buildrel W \over D}$ is defined by ${\buildrel W \over D} g =
a\otimes g$ with a 1--form $a$. This connection is related to the above one by
${\buildrel {\rm o} \over f} = 0$ and $a = - p = - g(\widetilde a, \cdot)$, that is, 
${\buildrel W \over D} = {\buildrel {\{\}} \over D} - \frac{1}{2} (\delta \otimes a + a
\otimes \delta - \widetilde a \otimes g )$. Because of the fact that $D$ is independent of conformal transformations, $a$ has
to transform according to $a
\mapsto a^\prime = a - d\lambda$. The geodesic equation reads ${\buildrel W \over D}_v v
= \beta v$. Manifolds endowed with a conformal structure and an equivalence class of
connections represented by the Weyl connection will be called a
{\it Weylian structure}. A space--time with the connection ${\buildrel W \over D}$ is
called a
{\it Weylian space--time} (For more interpretation see Perlick (1987)).

However, a violation of this compatibility condition does appear not only as an acausal
motion of point particles but also shows up in low velocity effects which we are going to
describe below. Consequently, with the help of these effects we can state a number of
alternative compatibility conditions which may have the advantage of being experimentally
more accessible than the causality condition of EPS. We see by inspection of
(\ref{geodesic3}) that
\begin{equation}
[D] = [{\buildrel W \over D} + {\buildrel {\rm o} \over f}] \, , \label{DDW}
\end{equation}
so that a Weylian structure is characterized by ${\buildrel
{\rm o} \over f} = 0$. Therefore any condition which also gives this result is equivalent
to the compatibility condition of EPS.

In the following we consider a space--time which consists of a manifold ${\cal M}$
endowed with a conformal and a projective structure which we do not assume to be
compatible.

\section{Measuring velocity, acceleration, and direction}

In this section we introduce that measurable quantities which are used in the following in order to state alternative versions of the compatibility axiom.

We want to describe the observation of the motion of a particle crossing the worldline of an observer.
In doing so neither the motion of the particle nor the motion of the observer are restricted in any way except that the observer is time--like with respect to the conformal structure.
All considerations in this chapter are carried through without use of the projective structure given by the geodesic equation (\ref{geodesic}); instead, we use (\ref{pathstructure}) only.
With these observable quantities it is possible to characterize the compatibility of the projective with the conformal structure in a way alternative to the EPS axiomatics.
(The following consideration may also be used for a low velocity characterisation of the projective structure.)

The observations will be carried through by means of a steady flow of light rays emitted from the observer to the particle which reflects these light rays.
The dates of emission and receipt of the rays are recorded by some clock.
This clock is just some monotonically growing parameter (see Fig.1).

In order to formalize this kind of observations, we at first consider a single observer described by a path $O: \mathbb{R} \rightarrow {\cal M}$ with some strongly monotonically growing parameter $t$.
Following Perlick (1987) we introduce with respect to some event $p_0 = O(t_0)$ a time-- and a distance function $\tau(q)$ and $\rho(q)$ provided $q$ lies within a sufficient small neighbourhood of $p_0$.
There is exactly one function  $\tau(q)$ and one function $\rho(q)$ so that the point $q$ can be connected by means of light rays with $O(\tau(q) + \rho(q))$ and $O(\tau(q) - \rho(q))$.

Next we describe along the lines of Perlick (1987) the observations made with a second particle $P: \mathbb{R} \rightarrow {\cal M}$ with parameter $t^\prime$.
We assume that $P$ crosses $O$ at $p_0 = O(t_0)$.
Then, by means of Einstein's synchronisation procedure, we can introduce a parameter on $P$ via $\tau(P(t)) = t$.
This means that $P(t)$ and $O(t)$ are synchroneous.
$x(t) := \rho(P(t))$ is the radial distance between $O(t)$ and $P(t)$.

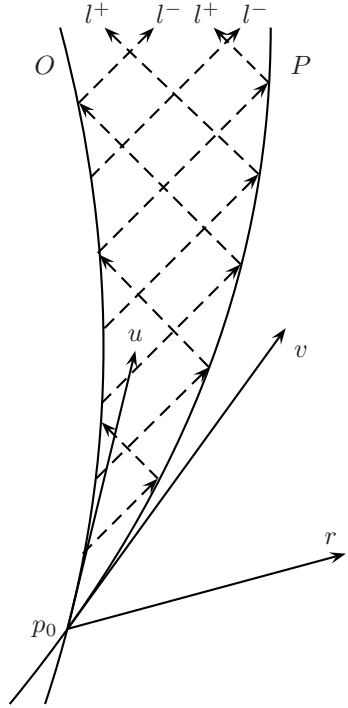
\begin{figure}[t]
\psset{unit=1cm}
\begin{center}
\begin{pspicture}(-2,-1)(6,8)%\showgrid
\psbezier(-0.47,-1)(2,2)(3,5)(3,8)
\psbezier(0,-1)(1,2)(1,5)(0.2,8)
\rput(0,7.5){$O$}
\rput(3.4,7.5){$P$}
\rput(0,0){$p_0$}
\psline{->}(0.3,0)(4,1)
\rput(3.8,1.2){$r$}
\psline{->}(0.3,0)(3.2,4)
\psline{->}(0.3,0)(1.2,3.7)
\rput(1.2,3.9){$u$}
\rput(3.4,3.7){$v$}
\psline[linestyle=dashed]{->}(0.5,1)(1.5,2)
\psline[linestyle=dashed]{->}(1.5,2)(0.75,2.75)
\psline[linestyle=dashed]{->}(0.68,2)(2.18,3.48)
\psline[linestyle=dashed]{->}(2.18,3.48)(0.72,4.98)
\psline[linestyle=dashed]{->}(0.75,3)(2.6,4.85)
\psline[linestyle=dashed]{->}(2.6,4.85)(0.45,7)
\psline[linestyle=dashed]{->}(0.79,4)(2.84,6.05)
\psline[linestyle=dashed]{->}(2.84,6.05)(0.79,8)
\psline[linestyle=dashed]{->}(0.7,5)(2.97,7.27)
\psline[linestyle=dashed]{->}(2.97,7.27)(2.24,8)
\psline[linestyle=dashed]{->}(0.6,6)(2.6,8)
\psline[linestyle=dashed]{->}(0.45,7)(1.45,8)
\rput(0.7,8.2){$l^+$}
\rput(1.65,8.2){$l^-$}
\rput(2.15,8.2){$l^+$}
\rput(2.8,8.2){$l^-$}
\end{pspicture}
\end{center}
\caption{Measurement of the elapsed time between emission and reception of light rays. $l^-$ are the emitted, $l^+$ the received light rays. $u$ and $v$ are the tangents to the worldlines $O$ and $P$, respectively.}
\end{figure}

The vector $u$ is the tangent to $O$ and $v$ the tangent to $P$.
The observation is made with light rays which are emitted from $O$ and received and reflected by $P$ and again received by $O$.
Thus the entities observed by $O$ are the direction of the emitted and received light rays $l^-$ and $l^+$ and their times of emission and reception.
The situation is illustrated by Fig.1.
This will be done in a small neighbourhood of the intersection point $p_0$ so that emitted (received) light rays do not intersect.
The set of emitted light rays $l^-$ define a two--surface $N_-$ and the set of received light rays $l^+$ a two--surface $N_+$.
These two surfaces can be parametrized by the time function $\tau$ and the distance function $\rho$ which define vector fields $\displaystyle R := {\partial\over{\partial\rho}}$ and $\displaystyle T := {\partial\over{\partial\tau}}$.
We have $\displaystyle \lim_{\rho\rightarrow 0}{T} = u$ and $l^+ \sim T - R$ and $l^- \sim T + R$.
Therefore
\begin{eqnarray}
g(T \pm R, T \pm R) & = & 0\, , \label{null}\\
{\buildrel {\{\}} \over D}_{T \pm R} (T \pm R) & = & \beta^\mp (T \pm R)\, .\label{null-geodesics}
\end{eqnarray}
In addition, every vector field $w$ on $N_+$ or $N_-$ may be decomposed with respect to $R$ and $T$ according to
\begin{equation}
w = w(\rho) R + w(\tau) T \, .\label{decomposition}
\end{equation}
If we take $\displaystyle v = {{dP(t)}\over{dt}}\in TN_\pm$, then we get $v(t)
= \dot x(t) R_{P(t)} + T_{P(t)}$. $\dot x(t)$ is the {\it radial velocity} of the passing
particle $P$ as observed by $O$. In $p_0$
\begin{equation}
v(t_0) = \dot x(t_0) r + u(t_0)
\end{equation}
with $\displaystyle r := \lim_{t\rightarrow t_0}{R_{P(t)}}$ and $g(r, r) = - g(u, u)$.
The velocity $x_0 := \dot x(t_0)$ is independent of the chosen parametrisation. If $v$
is tangent to a light ray, then $0 = g(v, v) = g(u + \dot x_0 r, u + \dot x_0
r) = g(u, u)\left( 1 -
{\dot x_0}^2\right)$, that is, $\dot x_0 = \pm 1$ (+1 for an outgoing, -1 for an incoming
light ray).

Equations (\ref{null}, \ref{null-geodesics} and \ref{decomposition}) are the mathematical
tools we need for the following where we are going to describe some notions which are
based upon the preceding considerations. For doing so we merely assume that the particle
paths are given by some particle structure (\ref{pathstructure}). By means of the
following procedures we can in principle search for a non--vanishing $\hbox{\teneufb f}$.

\subsection{The radial acceleration}

We want to determine the {\it radial acceleration} $\ddot x(t)$ of the crossing particle
$P$ as  measured by the observer $O$ with the parametrisation $t$. With the help of
(\ref{decomposition}) we can calculate
\begin{equation}
{\buildrel {\{\}} \over D}_v v = v(v(\rho))R + v(\rho)\left(v(\rho){\buildrel {\{\}} \over D}_R R + v(\tau) {\buildrel {\{\}} \over D}_T R \right) + v(v(\tau)) T + v(\tau) \left( v(\rho){\buildrel {\{\}} \over D}_R T + v(\tau) {\buildrel {\{\}} \over D}_T T \right)\, .
\end{equation}
By means of (\ref{null-geodesics}) in the form of $\beta^\pm l^\pm = D_{l^\pm} l^\pm = {\buildrel {\{\}} \over D}_T T \mp {\buildrel {\{\}} \over D}_T R \mp {\buildrel {\{\}} \over D}_R T + {\buildrel {\{\}} \over D}_R R$ we can eliminate ${\buildrel {\{\}} \over D}_R R$:
\begin{eqnarray}
{\buildrel {\{\}} \over D}_v v & = & \left[(v(v(\rho)) \mp \beta^\pm (v(\rho))^2\right] R + \left[v(v(\tau)) + \beta^\pm (v(\rho))^2\right] T \nonumber \\
& & \quad + \left[(v(\tau))^2 - (v(\rho))^2\right] {\buildrel {\{\}} \over D}_T T + 2 v(\rho)(v(\tau) \pm v(\rho)){\buildrel {\{\}} \over D}_T R \label{Dvv}
\end{eqnarray}
where, in addition, we used $[T, R] = 0$.
For $t \rightarrow t_0$ we get
\begin{equation}
{\buildrel {\{\}} \over D}_v v = (\ddot x_0 \mp \beta^\pm \dot x_0^2) R + \beta^\pm \dot x_0^2 T + (1 - \dot x_0^2){\buildrel {\{\}} \over D}_u u + 2 \dot x_0 (1 \pm \dot x_0){\buildrel {\{\}} \over D}_u r\, .
\end{equation}
Projecting this equation onto $R$ and $u$, subtracting (adding) the results and using $g({\buildrel {\{\}} \over D}_u u, u \mp r) = \pm g({\buildrel {\{\}} \over D}_u R, u \mp r)$ yields
\begin{equation}
g({\buildrel {\{\}} \over D}_v v, u \mp r) = \pm \ddot x_0 g(u, u) + (1 \pm  \dot x_0)^2
g({\buildrel {\{\}} \over D}_u u, u \mp r)\; . \label{accel1}
\end{equation}
Now we insert the general path structure (\ref{pathstructure}) and get
\begin{equation}
g(\hbox{\teneufb D}_v v - \hbox{\teneufb f}(v), u \mp r) = €\pm \ddot x_0 g(u, u) + (1
\pm
\dot x_0)^2 g(\hbox{\teneufb D}_u u - \hbox{\teneufb f}(u), u \mp r)\, .
\end{equation}
For freely falling particles $P$, that ist, for $\hbox{\teneufb D}_v v = \gamma v$, this amounts to
\begin{equation}
g(\gamma v - \hbox{\teneufb f}(v), u \mp r) = \pm \ddot x_0 g(u, u) + (1 \pm \dot x_0)^2
g(\hbox{\teneufb D}_u u - \hbox{\teneufb f}(u), u \mp r)\; . \label{TwoEqns}
\end{equation}
(We do not assume at the moment that $O$ is freely falling too, that is, for $u$ there may appear a force $\hbox{\teneufb F}$ in the equation of motion: $\hbox{\teneufb D}_u u = \alpha u + \hbox{\teneufb F}$.)
The unknown function $\gamma$ can be eliminated by considering the two equations contained in (\ref{TwoEqns}).
We get for the radial acceleration in the point $p_0$
\begin{equation}
{{\ddot x}\over{1 - \dot x_0^2}} = {1\over{g(u,u)}}\left( g(\hbox{\teneufb D}_u u - \hbox{\teneufb f}(u), r) - \dot x_0 g(\hbox{\teneufb D}_u u - \hbox{\teneufb f}(u), u) + {{g(f(v), r + \dot x_0 u)}\over{1 - \dot x_0^2}}\right)\, . \label{measured_accel}
\end{equation}
The observed acceleration depends on the velocity of the observed particle as well as on the form of the force $f(v)$ and on the state of motion and the parametrisation of the observer $O$.

If the path $P$ is a light ray, then we have to insert ${\buildrel {\{\}} \over D}_v v
\sim v$ with $g(v, v) = 0$ into (\ref{accel1}) and proceed as above. We get $\ddot x_0 =
0$. Consequently, light rays are always observed to be not accelerated irrespective of
the state of motion of the observer.

\subsection{A geometrical Coriolis acceleration}

In Galileian mechanics, the Coriolis force bends the path of some particle with respect to a non--inertial system if the particle passes the origin of this system.
By decomposing the coordinates of the particles into a radial and an angular part this results in a non--vanishing rotational motion of the position vector.
Similarly in our case a rotational motion of the same kind can be observed by comparing the direction of the incoming or the outgoing light rays $l^\pm$ with the direction defined by a bouncing photon (Pirani (1965)).
This amounts to consider the transport of the vector $R$ along the observers worldline $O(t)$ at the point $p_0$.
(Although the notion `Coriolis force' usually is taken for the bending of a path in a non--inertial system, we keep this notion, indicating that in our case where the inertial system is replaced by a non--rotating system in the sense of the bouncing photon, such a bending may still occur.
Here this bending is due to some geometrical field.)

For doing so we use Eq.(\ref{Dvv}) and solve for ${\buildrel {\{\}} \over D}_T R$. In the
following we use the projection operator $\displaystyle P_S := \delta - {{S g(S, \cdot)
}\over{g(S, S)}}$ for a non--null $S \in T{\cal M}$ which projects into the rest space of
$S$. The observable quantity describing the rotation of the direction $R$ is $P_R P_T
{\buildrel
{\{\}} \over D}_T R$ which will be taken at $p_0$ (we denote ${\buildrel {\{\}} \over
D}_u r = {\buildrel {\{\}} \over D}_T R$ at the point $p_0$):
\begin{eqnarray}
P_r P_u {\buildrel {\{\}} \over D}_u r & = & P_{r} P_u {1\over{2\dot x_0}}
\left({1\over{1 + \dot x_0}} {\buildrel {\{\}} \over D}_v v - (1 - \dot x_0){\buildrel
{\{\}} \over D}_u u\right) \nonumber \\ & = & P_{r} P_u {1\over{2\dot x_0}} \left({1\over{1
+ \dot x_0}}
\left(\hbox{\teneufb D}_v v - \hbox{\teneufb f}(v) \right) - (1 - \dot
x_0)\left(\hbox{\teneufb D}_u u - \hbox{\teneufb f}(u)\right)\right) \, .
\label{Cor_accel}
\end{eqnarray}
This relation is valid only for particles with $\dot x_0 \neq 0$. This transport equation
is independent of the parametrisation of the observer and therefore describes an
invariant measurable quantity. Because of $P_{P_T l^+}P_T {\buildrel {\{\}} \over D}_T
P_T l^+ = P_r P_u {\buildrel {\{\}} \over D}_u r$ in $p_0$ the corresponding observation
is most easily carried through by recording the directions of the incoming light rays
emitted by the particle $P$.

If $P$ is a light ray then again we get no effect: $P_r P_u {\buildrel {\{\}} \over D}_u r = 0$.

\subsection{The zig--zag construction}

In analogy to the bouncing photon construction of Pirani (1965) one can perform a 'bouncing particle', or a zig--zag construction, Ehlers and Schild (1973).
Although this could be done in a purely affine way we will use the metric instead.
The zig--zag construction consists in a central particle $O$ with tangent $u$ and other particles which, within some neighbourhood of the central particle, cross the central particle again and again after being reflected nearby, see Fig.2.
In this way the particles define a 2--surface, or, after projection into the rest space of the central particle, a direction in the rest space propagating along the path of the central particle.

\begin{figure}[t]
\begin{center}
\psset{unit=0.7cm}
\begin{pspicture}(0,-1)(10,11)%\showgrid
\psbezier(2.5,0)(4,4)(3,8)(3.7,12)
\psbezier[linewidth=1pt](3.8,-0.5)(6.2,4)(4,8)(6,12)
\psbezier(6,0)(7.4,4)(6,8)(7.7,12)
\rput(4,-0.9){\txt{central\\ particle}}
\rput(2,-0.3){\txt{auxilary\\ particle}}
\rput(6.2,-0.3){\txt{auxilary\\ particle}}
\psline[linestyle=dashed]{->}(4.3,0.6)(6.6,2.9)
\psline[linestyle=dashed]{->}(6.6,2.9)(3.4,6.1)
\psline[linestyle=dashed]{->}(3.4,6.1)(7,9.7)
\psline[linestyle=dashed]{->}(7,9.7)(4.7,12)
\psline[linestyle=dashed]{->}(4.3,0.6)(3,1.9)
\psline[linestyle=dashed]{->}(3,1.9)(6.7,5.6)
\psline[linestyle=dashed]{->}(6.7,5.6)(3.4,8.9)
\psline[linestyle=dashed]{->}(3.4,8.9)(6.5,12)
\psline[linewidth=1pt]{->}(4.3,0.6)(6,4.6)
\rput(6.2,4.5){$u$}
\rput(5.5,2.2){$l_1$}
\rput(3.7,1.7){$l_2$}
\end{pspicture}
\end{center}
\caption{Zig-zag construction around a central particle $O$.}
\end{figure}
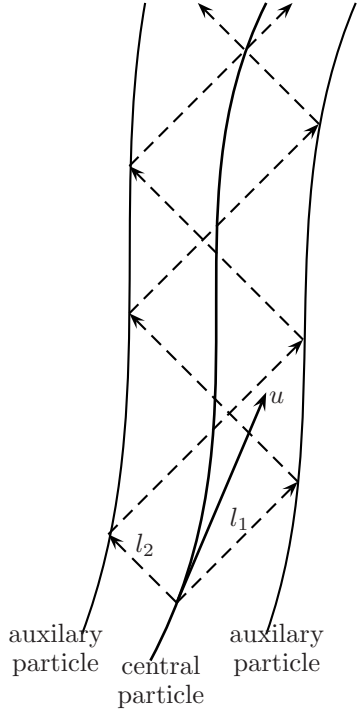

Again we make use of the general equation of motion (\ref{pathstructure}) for all particles.
The condition that the particles with tangent $v$ and $w$ lie in the same plane with the central particle $u$ is
\begin{equation}
u = \sigma v + \mu w \qquad\hbox{for some functions}\; \sigma, \mu \, .\label{uvw}
\end{equation}
The condition that, after reflection, the particles will cross the central worldline again, is secured by
\begin{equation}
{\cal L}_v w = \epsilon w + \delta v \qquad\hbox{for some functions}\; \epsilon, \delta  \, . \label{refluvw}
\end{equation}
We now want to determine the equation which governs the transport of the spacelike directions
\begin{equation}
V := P_u v \quad\hbox{or}\quad W := P_u w \; .\label{projectionoperator}
\end{equation}

With the help of $\sigma V = - \mu W$ we can calculate
\begin{equation}
P_V P_u {\buildrel {\{\}} \over D}_u V = {1\over\sigma}P_V P_u {\buildrel {\{\}} \over D}_u(\sigma V) = {1\over{2\sigma}}P_V P_u {\buildrel {\{\}} \over D}_u (\sigma V - \mu W).
\end{equation}
Inserting (\ref{uvw}, \ref{refluvw} and \ref{projectionoperator}) and the equations of motion (\ref{pathstructure}) for the particles $v$ and $w$, we finally get
\begin{equation}
P_V P_u {\buildrel {\{\}} \over D}_u V = {1\over{2\sigma}}P_V P_u \left( \mu^2 \hbox{\teneufb f}(w) - \sigma^2 \hbox{\teneufb f}(v) + \left(\mu^2 g(w,w) - \sigma^2 g(v,v)\right){{\hbox{\teneufb D}_u u - \hbox{\teneufb f}(u)}\over{g(u,u)}} \right) \label{bouncingparticle}
\end{equation}
This equation is invariant against reparametrisation of the paths, so that it indeed describes the propagation of the direction $V$.

If $v$ and $w$ are tangent to light rays, then the above equation reduces to
\begin{equation}
P_V P_u {\buildrel {\{\}} \over D}_u V = 0 . \label{bouncingphoton}
\end{equation}
This is precisely the characterisation of the bouncing photon usually taken as definition
for non--rotation, Pirani (1965). If for another space--like vector $q$ defined along $P$
the above expression  does not vanish, then the operator $\Omega$, which is a
$\left(1\atop 1\right)$--tensor defined by $P_q P_u {\buildrel {\{\}} \over D}_u q =
\Omega q$ along $P$, is called the {\it rotation} of $q$.

\section{Low velocity compatibility conditions}

In this section we show that certain demands on the radial acceleration, on the Coriolis acceleration of freely falling particles or on the zig--zag construction which are measured by an observer, who may be in arbitrary motion, can be fulfilled only if the equations of motions for the freely falling particles are restricted in such a way that space--time is described by a Weylian structure.

For doing so we assume a projective structure, which characterizes the strong equivalence principle for point particles.
In this case we get for a vector $v$ the decomposition $f(v) = f(v,v) =  {\buildrel {\rm o} \over f}(v,v) + 2 p(v) v + \widetilde a g(v, v)$.

\subsection{The compatibility with the standard clock}

For the projective structure represented by (\ref{geodesic3}) the measured radial acceleration (\ref{measured_accel}) results in
\begin{eqnarray}
{{\ddot x_0}\over{1 - {\dot x_0}^2}} & = & {{g(D_u u, r)}\over{g(u, u)}} - \dot x_0 \left({{g(D_u u, u)}\over{g(u, u)}} + {{{\buildrel {\rm o} \over f}(r,r,u) - {\buildrel {\rm s} \over f}(u,u,u)}\over{g(u, u)}} - g(\widetilde a, u) - p(u)\right) \nonumber\\
& & + {{\dot x_0}\over{g(u, u)(1 - {\dot x_0}^2)}}\Bigl({\buildrel {\rm s} \over f}(u, u,
u) + 3{\buildrel {\rm s} \over f}(u, r, r)) + \dot x_0 ({\buildrel {\rm s} \over
f}(r,r,r) + 3 {\buildrel {\rm s} \over f}(u,u,r)) \Bigr) \label{Proj_accel}
\end{eqnarray}
According to this equation the measured radial acceleration depends on the non-Weylian terms only in combination with the measured velocity.
Therefore, having in mind the definition of a standard clock of Perlick (1987), we can use this observation to state:

\bigskip
\noindent {\bf Theorem:} {\it Space--time has a Weylian structure iff for all observers there exists a parameter so that for all freely falling particles the observable quantity $\displaystyle {{\ddot x_0}\over{1 - {\dot x_0}^2}}$ is independent of the velocity $\dot x_0$ for $|\dot x_0| \ll 1$.}

\medskip
{\it Proof:}
Since the ''only--if'' part of the proof is clear we proof the ''if''--part only.
Expansion of the right--hand--side of (\ref{Proj_accel}) up to third order in $\dot x_0$
and the assumption that the result is independent of $\dot x_0$ gives three conditions
\begin{eqnarray}
0 & = & {\buildrel {\rm s} \over f}(u, u, u) + 3 {\buildrel {\rm s} \over f}(u,r,r) \label{cond1} \\
0 & = & {\buildrel {\rm s} \over f}(r, r, r) + 3 {\buildrel {\rm s} \over f}(u, u, r)\label{cond2} \\
0 & = & {{g(D_u u, u)}\over{g(u, u)}} + {{{\buildrel {\rm o} \over f}(r,r,u) - {\buildrel {\rm s} \over f}(u,u,u)}\over{g(u, u)}} - g(\widetilde a, u) - p(u)  \label{cond3}
\end{eqnarray}
which are valid for all $r$ and $u$ with $g(u, r) = 0$.
Addition of (\ref{cond1}) and (\ref{cond2}) gives $0 = {\buildrel {\rm s} \over f}(u + r, u + r, u + r) \quad \forall \; \hbox{timelike} \; u, \forall \; \hbox{spacelike} \; r$ with $g(r, r) = - g(u, u)$ and $g(u, r) = 0$.
This is equivalent to
\begin{equation}
0 = {\buildrel {\rm s} \over f}(l, l, l) \qquad \forall l \;\hbox{with}\; g(l, l) = 0 \,
.
\end{equation}
From this equation we conclude ${\buildrel {\rm s} \over f} = 0$.
Insertion of this result into  (\ref{cond3}) gives
\begin{equation}
0 = g({\buildrel W \over D}_u u, u) + {\buildrel {\rm o} \over f}(r, r, u) \qquad \forall \; \hbox{timelike} \; u, \; \hbox{spacelike} \;r
\end{equation}
where we used ${\buildrel W \over D}$ defined at the end of section 2. This implies
\begin{equation}
{\buildrel {\rm o} \over f}(r, r, u) = {\buildrel {\rm o} \over f}(r^\prime, r^\prime, u) \qquad \forall \; \hbox{spacelike} \; r, r^\prime\; . \label{xxx}
\end{equation}
Now we take another particle with $r^\prime = r + \delta r$ and small $\delta r$. Because
of $g(r,r) = g(r^\prime, r^\prime) = - g(u,u)$ we have to first order in $\delta r$:
$g(u, \delta r) = g(r, \delta r) = 0$. Insertion into (\ref{xxx}) gives $P_u P_r
{\buildrel {\rm o} \over f}(r, \cdot, u) = 0, \; \forall \; \hbox{spacelike} \; r,
\forall \; \hbox{timelike} \; u$. After some calculation this condition yields ${\buildrel
{\rm o} \over f} = 0$. The final condition $0 = g({\buildrel W \over D}_u u, u)$
merely singles out a distinguished parametrization of the observer's path but gives no
further constraint on the geometry. \hfill$\diamond$

\bigskip
Therefore, for an arbitrary parametrization, we finally get as measured radial acceleration in a Weylian space--time
\begin{equation}
{{\ddot x_0}\over{1 - {\dot x_0}^2}} = {{g({\buildrel W \over D}_u u, r)}\over{g(u, u)}} - \dot x_0 {{g({\buildrel W \over D}_u u, u)}\over{g(u, u)}} \label{accelWeyl}
\end{equation}
which has already been calculated by Perlick (1987) assuming a Weyl geometry from the
beginning. The term $g({\buildrel W \over D}_u u, u)$ on the right hand side of
(\ref{accelWeyl}) is responsible for the acceleration to depend on the parameter chosen
on the wordline of the observer. Because this term is connected with the measured
velocity one can,
according to Perlick (1987), define a {\it standard clock} as that parameter $t$ of an observer for which the left hand side of (\ref{accelWeyl}) %quantity $\displaystyle {{\ddot x}\over{1 - {\dot x}^2}}$
is independent of $\dot x_0$. Therefore we can state:

\bigskip
\noindent {\bf Corollary:} {\it A standard clock only exists in a Weylian structure.}

\bigskip
The reason for the fact that one cannot define a standard clock in a non--Weylian structure can be read off from Eq.(\ref{Proj_accel}): The accereleration not only depends on the radial velocity $\dot x$ but also on the direction $r$ along which the particle $P$ propagates.
If one is given a transport law for directions $r$ (e.g. (\ref{Proj_Coriolis}) or (\ref{Proj_zigzag}), see below) then it is possible to define for each direction $r$ a distinguished parameter similar to the standard clock of Perlick.
However, in this case a clock is not a geometrically induced property of a {\it path}; it also depends on an additional rest--space direction.
Of course, this feature may be regarded as tool to measure the non--Weylian parts of the connection:
The differences of the time intervals for the various clocks which depend on directions $r$ are directly connected with the non--Weylian parts ${\buildrel {\rm o} \over f}$.

The above corollary means that non--Weylian parts of a connection are obstructions for
the construction of a unique standard clock. Note that the above procedure shows that the
equation of motion for the observer has to be written in terms of the Weyl connection
${\buildrel W \over D}$: An observer $O$ with a standard clock is described uniqueley by
the equation ${\buildrel W \over D}_v v = \hbox{\teneufb f}$ where $\hbox{\teneufb f}$ is
a force acting on $O$ in its rest space: $g(\hbox{\teneufb f}, u) = 0$.

\subsection{Geometrical Coriolis acceleration}

For the projective structure given by (\ref{geodesic3}) the measured Coriolis acceleration (\ref{Cor_accel}) is ($\dot x \neq 0$)
\begin{equation}
P_r P_u {\buildrel {\{\}} \over D}_u r = - {1\over{2 ( 1 + \dot x_0)}} P_r P_u\left( 2
{\buildrel {\rm o} \over f}(u, r) + \dot x_0 ({\buildrel {\rm o} \over f}(u,u) + {\buildrel
{\rm o} \over f}(r, r))\right) - {{1 - \dot x_0}\over{2\dot x_0}} P_r P_u D_u u \, .
\label{Proj_Coriolis}
\end{equation}
In the case of a nongeodesic observer this acceleration becomes larger for smaller velocities $\dot x_0$ because the curvature of the path in the rest space of $u$ becomes larger.

Since the Coriolis acceleration is connected with the non--Weylian part ${\buildrel {\rm o} \over f}$ we can state the

\bigskip
\noindent {\bf Theorem:} {\it Space--time has a Weylian structure iff $\dot x_0 (1 + \dot x_0) P_r P_u {\buildrel {\{\}} \over D}_u r$ does not depend on $\dot x_0$ for small $\dot x_0$.}

\medskip
{\it Proof:} Expansion of $\dot x_0 (1 + \dot x_0) P_r P_u {\buildrel {\{\}} \over D}_u r$ with respect to $\dot x_0$ leads to the condition $P_u P_r {\buildrel {\rm o} \over f}(u, r) = 0$ for all timelike $u$ and spacelike $r$ with $g(u, r) = 0$ which gives ${\buildrel {\rm o} \over f} = 0$. \hfill$\diamond$

\bigskip
Special cases are provided by the next two corollaries.

\bigskip
\noindent {\bf Corollary:} {\it Space--time has a Weylian structure iff the Coriolis acceleration of all freely falling particles measured by freely falling observers does not depend on the velocity $\dot x_0$ for small $\dot x_0$.}

\medskip
{\it Proof:} The condition that the right hand side of (\ref{Proj_Coriolis}) should be independent of $\dot x_0$ for small $\dot x_0$ implies $2 {\buildrel {\rm o} \over f}(u, r) = {\buildrel {\rm o} \over f}(u,u) + {\buildrel {\rm o} \over f}(r, r)$ for all spacelike $r$ and timelike $u$.
This is equivalent to ${\buildrel {\rm o} \over f}(u - r, u - r) = 0$, or ${\buildrel {\rm o} \over f}(l, l) = 0 \;\;\forall l$ with $g(l,l) = 0$.
Consequently, ${\buildrel {\rm o} \over f} = 0$ since ${\buildrel {\rm o} \over f}$ is tracefree. \hfill$\diamond$

\bigskip
\noindent {\bf Corollary:} {\it The Coriolis acceleration of all freely falling particles measured by freely falling observers vanishes only if space--time has a Weyl geometry.}

\subsection{The zig--zag construction and the bouncing photon}

For the projective structure (\ref{geodesic3}) the zig--zag construction (\ref{bouncingparticle}) gives
\begin{equation}
P_V P_u {\buildrel {\{\}} \over D}_u V = - P_V P_u \left({\buildrel {\rm o} \over f}(V,u) - {1\over{2\sigma g(u,u)}} (\sigma^2 g(v,v) - \mu^2 g(w,w)) D_u u \right) \label{Proj_zigzag}
\end{equation}
Again, ${\buildrel {\rm o} \over f}$ disturbs distinguished transportation features.
Therefore we can state (the proof is the same as above):

\bigskip
\noindent {\bf Theorem:} {\it Space--time has a Weylian structure iff the vector $V$ constructed according to the zig--zag construction with freely falling central particle does not rotate.}

\bigskip
We can also make two zig--zag constructions and compare the motion of the two space--like
directions which are defined by these two constructions. For this we introduce as usual
the notion of the angle $\varphi$ between two space--like vectors $V$ and $V^\prime$
through $\displaystyle \cos\varphi = {{- g(V, V^\prime)}\over\sqrt{g(V, V)\, g(V^\prime,
V^\prime)}}$. Then we can state the

\bigskip
\noindent {\bf Theorem:} {\it Space--time has a Weylian structure iff the angle between two vectors $V$ and $V^\prime$ constructed according to two zig--zag construction is constant.}

\bigskip
This statement means that only in a Weyl structure there is a kind of isotropy: for each freely falling central particle no direction in its rest space can be distinguished by means of the zig--zag constructuon.

\section{Discussion}

We have shown that it is possible to replace the compatibility axiom of EPS which needs massive point particles with velocities arbitrarily close to the velocity of light, by a number of requirements using point particles with low velocity only.
These alternative compatibility conditions require the velocity--independence of the radial acceleration, the vanishing of the Coriolis acceleration and of the zig-zag construction.
The latter two can (in principle) be tested by null experiments.

A next step might be to apply these low velocity compatibilty conditions to the general path structure.
Having the analysis of Coleman and Korte (1984) in mind one expects that even in this genaralized situation the Weylian structure will be the only geometrical frame within which the outcome of the respective experiments is as demanded.

Another step in a space--time axiomatics consists in the reduction of the Weyl structure to a Riemannian one.
This has to be done by calculating specific effects connected with the Weylian field strength $f := da$.
Effects of such type are the second clock effect or an additional Weyl induced rotation of congruences in Weylian space--times, see Perlick (1991).
The requirement that such effects are absent is sufficient for the intended reduction.
Another method for the reduction to a Riemannian structure by classical means is presented by Schelb (1997).
However, on a classical level such experimental constructions don't seem to be easily realisable.
Only the identification of standard clocks with atomic clocks or other quantum standards leads to the result that with high accuracy there is no second clock effect.
Indeed, it is a general feature that space--time axiomatics using elements of quantum mechanics from the very beginning automatically lead to a Riemannian structure of space--time (Audretsch and L\"ammerzahl (1993)).

\section*{Acknowledgement}

The author gratefully thanks V.\ Perlick for discussions and the referees for helpful comments.

\section*{References}

\begin{list}{}{\leftmargin40pt\itemindent-40pt\parsep-3pt}

\item Audretsch J., L\"ammerzahl C. (1993): A New Constructive
Axiomatic Scheme for the Geometry of Space--Time, {\bf in:} Majer U.,
Schmidt H.-J. (edts.): {\it Semantical Aspects of Space--Time
Geometry}, BI Verlag, Mannheim, p.21.

\item Coleman R.A., Korte H. (1980): Jet bundles and path structures, {\it J. Math. Phys.} {\bf 21} 1340.

\item Coleman R.A., Korte H. (1984): Constraints on the nature of inertial motion arising from the universality of free fall and the conformal causal structure of space--time, {\it J. Math. Phys.} {\bf 25} 3513.

\item Coleman R.A., Korte H. (1987): Any physical monopole equation of motion structure uniquely determines a projective structure and an $(n-1)$--force, {\it J. Math. Phys.} {\bf 28} 1492.

\item Coleman R.A., Korte H. (1993): Why fundamenal structures are first or second order, preprint., Zentrum f\"ur interdisziplin\"are Forschung, Universit\"at Bielefeld.

\item Coleman R.A., Schmidt H.-J. (1993): A geometric formulation of the equivalence principle, preprint, Zentrum f\"ur interdisziplin\"are Forschung, Universit\"at Bielefeld .

\item Douglas J. (1928): The general geometry of paths, {\it Ann. Math.} {\bf 29} 143.

\item Ehlers J., Pirani F.A.E., Schild A. (1972): The Geometry of Free Fall and Light Propagation, in: L. O'Raifeartaigh (ed.): {\it General Relativity, Papers in Honour of J.L. Synge}, Clarendon Press, Oxford.

\item Ehlers J., K\"ohler E. (1977): Path Structure on Manifolds, {\it J. Math. Phys.} {\bf 18} 2014.

\item Ehlers J., Schild, A. (1973): Geometry in a Manifold with Projective Structure, {\it Comm. Math. Phys.} {\bf 32} 119.

\item Heilig U., Pfister H. (1991): Characterization of free fall paths by a global or local Desargues property, {\it JPG} {\bf 7} 419.

\item Perlick V. (1987): Characterisation of Standard Clocks by Means of Light Rays and Freely Falling Particles, {\it Gen. Rel. Grav.} {\bf 19} 1059.

\item Perlick V. (1991): Observer fields in Weylian spacetime models, {\it Class. Quantum Grav.} {\bf 8} 1369.

\item Pirani F.A.E. (1965): A note on bouncing photons, {\it Bull. Acad. Polon. Sci., Ser. Sci. Math. Astr. Phys.} {\bf 13} 239.

\item Rund, H. (1959): {\it The Differential Geometry of Finsler Spaces}, Grundlehren der Mathematischen Wissenschaften in Einzeldarstellungen, Springer--Verlag, Ber\-lin.

\item Schelb U. (1992): An axiomatic basis of space--time theory, part III: Construction of a differentiable manifold, will apear in {\it Rep. Math. Phys.} {\bf 31} 297.

\item Schelb U. (1997): On a new condition distinguishing Weyl and Lorentz space--times, {\it Int. J. Theor. Phys.} {\bf 36} 1341.

\item Schr\"oter J. (1988): An axiomatic basis of space--time theory, part I: Construction of a causal space with coordinates, {\it Rep. Math. Phys.} {\bf 26} 305.

\item Schr\"oter J., Schelb U. (1992): An axiomatic basis of space--time theory, part II: Construction of a $C^0$--manifold, {\it Rep. Math. Phys.} {\bf 31} 5.

\item Schr\"oter and Schelb (1995): Remarks concerning the notion of free fall in axiomatic space--time theory, {\it Gen. Rel. Grav.} {\bf 27} 605.

\end{list}

\clearpage

\thispagestyle{empty}
\baselineskip14pt

\normalsize
\vspace*{5mm}
Dr. Claus L\"ammerzahl

\vspace*{1.2cm}
To \hfill July 13th, 1993

Prof. Dr. A. Held

Institute of Theoretical Physics

Room 121

Sidlerstr. 5

\vspace*{2mm}
CH - 3012 Berne

\vspace*{2mm}
Switzerland

\vskip 2cm
Dear Professor Held,

\vspace{7mm}
enclosed you will find 3 copies of my manuscript entitled

\medskip
A Characterisation of the Weylian Structure of Space-Time

by Means of Low Velocity Tests

\medskip
\noindent for publication in {\it General Relativity and Gravitation}.

\bigskip
\hspace*{4cm} Sincerely Yours

\vspace*{1.5cm}
\hspace*{7cm} (C. L\"ammerzahl)

\vspace*{1cm}
Phone: +49(0)7531 88-3795

Fax: +49(0)7531 88-3888

e-mail: Claus@spock.physik.uni-konstanz.de

%\newpage

%\begin{itemize}

%\item Wegestruktur, oder geodaetische Struktur eventuell besser mit 'direction field'?

%\item Equation(11): In general, we get as angle between $u$ and $v$ in $p_0$: $\displaystyle \cos\varphi = {1\over\sqrt{1 - {\dot x}^2}}$.

%\end{itemize}

\end{document}